\documentclass[pra,twocolumn,preprintnumbers,amsmath,amssymb,superscriptaddress,showpacs,longbibliography]{revtex4-2}
\usepackage[T1]{fontenc} 
\usepackage[utf8]{inputenc} 

\usepackage{graphicx}
\usepackage{latexsym}
\usepackage{amsmath}
\usepackage{graphics}
\usepackage{amssymb}
\usepackage{layout}
\usepackage{verbatim}
\usepackage{amsfonts,epsfig,dsfont}
\usepackage{soul}
\usepackage[dvipsnames,table]{xcolor} 
\usepackage{hyperref}
\usepackage{tikz}
\usepackage{tikz-feynman}
\usepackage{color}
\hypersetup{colorlinks, linkcolor={blue!90!black}, citecolor={blue!90!black}, urlcolor={blue!90!black} }

\usepackage[customcolors]{hf-tikz}

\newcommand{\tr}{\hbox{Tr}}
\usepackage{ulem} 

\begin{document}

\title{Convex-roof entanglement measures of density matrices block diagonal in disjoint subspaces
	for the study of thermal states
}
\author{Mikołaj Jędrzejewski}
\affiliation{Department of Theoretical Physics, Faculty of Fundamental Problems of Technology, Wroc{\l}aw University of Science and Technology,
	50-370 Wroc{\l}aw, Poland}
\author{Kacper Kinastowski}
\affiliation{Department of Theoretical Physics, Faculty of Fundamental Problems of Technology, Wroc{\l}aw University of Science and Technology,
	50-370 Wroc{\l}aw, Poland}
\author{Katarzyna Roszak}
\affiliation{Department of Optics, Palack{\'y} University, 17. Listopadu 12, 771 46 Olomouc, Czech Republic}
\affiliation{Department of Theoretical Physics, Faculty of Fundamental Problems of Technology, Wroc{\l}aw University of Science and Technology,
50-370 Wroc{\l}aw, Poland}

\date{\today}

\begin{abstract}
	We provide a proof that entanglement of any density matrix which block diagonal
	in subspaces which are disjoint in terms of the Hilbert space of one of the two potentially entangled subsystems can simply be calculated as the weighted average of entanglement present within each block.
	This is especially useful for thermal-equilibrium states which always inherit the symmetries present
	in the Hamiltonian, since block-diagonal Hamiltonians are common as are interactions which
	involve only a single degree of freedom of a greater system. We exemplify our method on a simple
	Hamiltonian, showing the diversity in possible temperature-dependencies of Gibbs state entanglement
	which can emerge in different parameter ranges.
\end{abstract}
\maketitle

\section{Introduction}

The quantification of entanglement for mixed states is a complicated problem
especially once the system under study becomes larger. This is reflected in the number of 
entanglement measures which have been proposed and which are used to date,
and the fact that no unique entanglement measure for mixed states has been established \cite{plenio07}.
Each comes with its own set of problems, 
which can be broadly classified as either 
the need of numerical optimization of complex functions
over a set of parameters which grows rapidly with system size,
or the inability to detect certain types of entanglement.
The first is characteristic for such measures as distillable entanglement \cite{bennett96,vedral97,rains99,eisert00}, entanglement cost \cite{hayden01}
and all convex-roof entanglement measures, such as Entanglement of Formation (EoF) \cite{bennett96a,bennett96}.
Negativity \cite{vidal02,lee00a,plenio05b} on the other hand is much easier to compute, since it only requires diagonalization, but it has no clear physical interpretation and
it suffers from the second drawback, since it cannot detect bound entangled states \cite{horodecki97b,horodecki98,smolin01,diguglielmo11,hiesmayr13,sentis18,gabodulin19}.
Bound entangled states can occur only for systems of dimension larger than $6$
\cite{horodecki00}, so they are a small problem unless systems are relatively large,
but they have been found already for very mixed states of two qutrits \cite{horodecki97b}.

The quest to find efficiently computable entanglement measures that have a clear
physical interpretation 
continues \cite{roszak20,wang20,bergh21}, because entanglement remains an important factor
which predetermines the possibility of very
quantum and nonintuitive behavior of bipartite systems. 
One approach is to sacrifice generality in order to obtain solvable scenarios 
which allows for the study of entanglement present in density matrices generated 
by a certain class of Hamiltonians. This has been done for qubits interacting with large
environments via a Hamiltonian which leads to qubit pure decoherence \cite{roszak15,roszak17,roszak20}
and then generalized to systems of any size \cite{roszak18}. The consequence here was that 
increased understanding of this type of entanglement allowed for schemes for its measurement
to be devised \cite{roszak19a,strzalka20,strzalka21},
which have been experimentally tested \cite{zhan21}.

We will be focusing on the class of convex roof entanglement measures, which are easiest
to understand on an intuitive level, as they are defined by an extension of any pure-state entanglement
measure to mixed states. They come with a seemingly straightforward instruction for computation,
as one is to calculate the mean entanglement of the state under
study decomposed into pure states 
and then minimize over all possible decompositions. There even exists a way to find
EoF directly from the density matrix for a system of two qubits \cite{wootters98}
and efficient minimization techniques for small bipartite systems \cite{audenaert01}.
Nevertheless, for larger systems these measures are usually not found directly using the instruction,
but are evaluated on the basis of various techniques used to make minimization efficient
or to bypass it entirely. These include approximations of the lower bounds of entanglement measures 
\cite{toth15,zhang16,carrijo21},
direct methods for states possessing additional symmetries \cite{terhal00b,giedke03,ryu12,regula16},
variational approaches to minimization and conjugate gradient methods \cite{audenaert01,rothlisberger09,ryu08}, and
the minimization over pure state extensions of the state \cite{allende15}.
Antother technique exchanges minimization for
random sampling of possible pure-state decompositions
\cite{zyczkowski99,akulin15,gabodulin19}.
The majority of the techniques benefit greatly
from a reduction of the size of the Hilbert space of the system under study.

In our study we focus on a class of bipartite density matrices which are block diagonal 
in subspaces of the Hilbert space which are disjoint in terms of states of one of the 
two subsystems. We show that for such density matrices
any convex-roof entanglement measure can be found as the average of the amount of entanglement
contained within each block. Hence the minimization has only to be performed 
within the disjoint subspaces separately, substantially reducing the complexity of entanglement
quantification. One consequence of this
from a quantum information perspective is that bound entanglement cannot transcend between such blocks.
This allows criteria which disqualify the possibility of the appearance of bound entanglement 
to be applied 
to the blocks separately, thus extending the applicability of rank and system size constraints
to larger systems. 

A relevant question is whether the desired structure can appear in physical systems.
We mainly focus here on thermal states due to renewed interest in the context 
of quantum information \cite{markham08,li11,weedbrook12,wu19,motta20}. Thermal states have the advantage that the symmetries within them are directly
transferred from the Hamiltonian, thus if a Hamiltonian is block-diagonal, the corresponding
thermal state will also be block diagonal within the same subspaces. 
The condition for the blocks to be disjoint in terms of one subsystem
is commutation of the full Hamiltonian with a nontrivial operator acting solely on this part
of the system.
We provide examples of such Hamiltonians stemming from solid state physics.
Incidentally, for this class of Hamiltonians also time evolution can retain
the block diagonal structure, but this requires additional restrictions to be imposed
on the initial state.

We provide exemplary dependencies of entanglement on temperature
in Gibbs states for block-diagonal Hamiltonians, where the disjoint blocks are small enough to allow 
for the use of Wootters formula \cite{wootters98}, while the whole system is very large.
This allows us to demonstrate the efficiency of the method and study interplay of entanglement
characteristics of each block within the whole state. We find that parameter changes in the
Hamiltonian can lead to qualitative changes in the behavior of entanglement, some 
dependent on how entanglement within a given block reacts to temperature-induced mixing of the state,
but most related to how the different blocks are mixed. Hence, the use of our simplified formula
allows us to draw conclusions about the physics of entanglement.

The paper is organized as follows. In Sec.~\ref{sec2} we describe the class of mixed states under study
and in Sec.~\ref{sec3} we briefly comment on convex-roof entanlement measures. Sec.~\ref{sec4}
contains the proof that calculation of such measures for block-diagonal density matrices of the
studied type can be reduced to averaging of entanglement within separate blocks.
In Sec.~\ref{sec5} we comment on the consequences of this fact from a quantum information
perspective, while in Sec.~\ref{sec6} we discuss the type of systems which will yield 
the desired form of the state at thermal equilibrium and during time-evolution.
In Sec.~\ref{sec6} we provide a simple example and study temperature-dependence
of entanglement in different parameter ranges. Sec.~\ref{sec7} concludes
the paper.

\section{Entanglement in density matrices block diagonal in disjoint subspaces \label{sec2}}

We are studying bipartite density matrices and for clarity we will be calling one part the quantum
system (QS)
and the other part the environment (E). This is because the density matrices under consideration
are asymmetric with respect
to the two subsystems and an easily made distinction is necessary. 

The first assumption made on the form of the joint QS-E density matrices
under study is that they are block diagonal
in some separable QS-E basis. Let us denote this basis as $\{|s\rangle\}$ for QS
and $\{|e\rangle\}$ for E, so the density matrix is block-diagonal in basis $\{|se\rangle\equiv|s\rangle\otimes|e\rangle\}$. We will keep the notation with QS states on the left and E states
on the right of the ket/tensor product throughout the paper. 
We further assume that the block-diagonal form is the result of the properties of E, so that 
each block is built of QS-E states in which the whole range QS states can be present, but 
the set of E states in each block is orthogonal to the states in any other block.
Hence the class of density matrices under study can be written as 
\begin{equation}
\label{rhon}
\hat{\rho}=\sum_np_n\hat{\rho}_n,
\end{equation}
where the index $n$ distinguishes between the blocks, $p_n$ are probabilities with $\sum_np_n=1$
and $\hat{\rho}_n$ are density matrices occupying disjoint blocks in terms of E subspaces.
These density matrices not only fulfill $\hat{\rho}_n\hat{\rho}_m=0$ for $n\neq m$, but each can be 
written as
\begin{equation}
\label{rhonn}
\hat{\rho}_n=\sum_{ss'}\sum_{e_ne'_n}c_{ss'}^{ee'}|s\rangle\langle s'|\otimes|e_n\rangle\langle e'_n|,
\end{equation}
where all states $|e_n\rangle$ for a given $n$ constitute a separate subspace of E, so that
$\langle e_n|e'_m\rangle=0$ for all $e$ and $e'$ if $n\neq m$.

Let us exemplify this on the simplest possible nontrivial density matrix, when QS is a qubit with states 
$|0\rangle$ and $|1\rangle$
while E is of dimension $N=4$ with states $\{|0\rangle,|1\rangle,|2\rangle,|3\rangle\}$.
A mixed QS-E state
\begin{equation}
\hat{\rho}=\frac{1}{4}\left(
|\Psi_{01}\rangle\langle\Psi_{01}|
+|\Phi_{01}\rangle\langle\Phi_{01}|
+|\Psi_{23}\rangle\langle\Psi_{23}|
+|\Phi_{23}\rangle\langle\Phi_{23}|
\right),
\end{equation}
with Bell like components
\begin{subequations}
	\label{bell}
\begin{eqnarray}
|\Psi_{ij}\rangle&=&\frac{1}{\sqrt{2}}\left(|0i\rangle+|1j\rangle\right),\\
|\Phi_{ij}\rangle&=&\frac{1}{\sqrt{2}}\left(|1i\rangle+|0j\rangle\right),
\end{eqnarray}
\end{subequations}
can be written in matrix form as
\begin{equation}
\label{macex}
\hat{\rho}=\frac{1}{8}\left(
\begin{array}{*{8}{c}}
\hfsetfillcolor{white}
\hfsetbordercolor{blue}
\tikzmarkin{a}(0.1,-0.1)(-0.1,0.35)
\hfsetfillcolor{white}
\hfsetbordercolor{red}\tikzmarkin{c}(0.1,-0.1)(-0.1,0.35)1&1&0&0&0&0&0&0\\
1&1\tikzmarkend{c}&0&0&0&0&0&0\\
0&0&
\hfsetfillcolor{white}
\hfsetbordercolor{red}\tikzmarkin{d}(0.1,-0.1)(-0.1,0.35)1&1&0&0&0&0\\
0&0&1&1\tikzmarkend{d}\tikzmarkend{a}&0&0&0&0\\
0&0&0&0&
\hfsetfillcolor{white}
\hfsetbordercolor{blue}\tikzmarkin{b}(0.1,-0.1)(-0.1,0.35)
\hfsetfillcolor{white}
\hfsetbordercolor{red}\tikzmarkin{e}(0.1,-0.1)(-0.1,0.35)1&1&0&0\\
0&0&0&0&1&1\tikzmarkend{e}&0&0\\
0&0&0&0&0&0&
\hfsetfillcolor{white}
\hfsetbordercolor{red}\tikzmarkin{f}(0.1,-0.1)(-0.1,0.35)1&1\\
0&0&0&0&0&0&1&1\tikzmarkend{f}\tikzmarkend{b}
\end{array}
\right).
\end{equation}
The QS-E basis here is separable and is arranged in the order $\{|00\rangle,|11\rangle,|01\rangle,|10\rangle,|02\rangle,|13\rangle,|03\rangle,|12\rangle\}$
to highlight the block diagonal form. 

Two block structures are marked on the matrix (\ref{macex}). The smaller blocks (marked in red)
each encompass a subspace of the Hilbert space characteristic for a single of the Bell-like states
(\ref{bell}). These blocks do not fulfill our requirements, since the same E states can occur
in different blocks, that is $\{|0\rangle,|1\rangle\}$ in the upper-left two blocks,
and $\{|2\rangle,|3\rangle\}$ in the lower-right blocks.
We will be studying density matrices where the block-diagonal form is of the type
marked by the larger (blue) blocks, which encompass different subsets of the Hilbert space of E.

There is a fundamental difference in these two types of blocks with respect to entanglement.
In the exemplary state, each smaller block encompasses a maximally entangled Bell-like state,
but the mixture of two states in each larger block is separable \cite{su11}, since 
\begin{eqnarray}
&&\frac{1}{2}\left(|\Psi_{ij}\rangle\langle\Psi_{ij}|
+|\Phi_{ij}\rangle\langle\Phi_{ij}|\right)\\
\nonumber
&&=
\frac{1}{2}\left(|+\rangle\langle +|\otimes|+\rangle\langle +|
+|-\rangle\langle -|\otimes|-\rangle\langle -|\right),
\end{eqnarray}
with $|\pm\rangle=1/\sqrt{2}\left(|0\rangle\pm|1\rangle \right)$ on QS
and $|\pm\rangle=1/\sqrt{2}\left(|i\rangle\pm|j\rangle \right)$ on E.
Entanglement of the whole density matrix (\ref{macex}) depends on the interplay of the states
contained in the small blocks, but it does not depend on the interplay of the states
encompassed by the large blocks, since the separable form can be obtained in each of them
separately.

As a central result of this paper, we will show that this result is general for convex-roof entanglement measures, and that
the entanglement for any density matrix which is block diagonal in disjoint subspaces (\ref{rhon})
is given by
\begin{equation}
\label{eog}
E(\hat{\rho})=\sum_np_nE(\hat{\rho}_n).
\end{equation} 
Here $E(...)$ denotes the convex-roof entanglement measure of choice.

\section{Convex-roof entanglement measures \label{sec3}}

We will study convex-roof entanglement measures \cite{vidal00,plenio07}
for bipartite mixed state entanglement. They constitute an extension of pure-state
entanglement measures to mixed states and the class of measures is defined as follows.
Given a good pure state entanglement measure such as the entropy 
(this can be von Neumann entropy, any of the Renyi entropies, or linear entropy)
of the reduced density matrix of one of the potentially entangled subsystems, $E(|\psi\rangle)$,
one can construct a mixed state measure. This is done by 
first providing a decomposition of the density matrix into pure states,
\begin{equation}
\label{rhodec}
\hat{\rho}=\sum_nP_n|\psi_n\rangle\langle\psi_n|,
\end{equation}
where $P_n$ are probabilities and the states $|\psi_n\rangle$ do not have to be orthogonal.
Note that there is an infinite number of such decompositions
if the state is not pure and a full parametrization of the decompositions of a given mixed
state quickly grows in complexity (and the number of free paramters) with the size of the system
\cite{nielsen00,gurvits03b,gurvits04}.
The average of pure-state entanglement
\begin{equation}
\label{eav}
\tilde{E}(\hat{\rho})=\sum_nP_nE(|\psi_n\rangle),
\end{equation}
is not enough to quantify entanglement of state (\ref{rhodec}), because
the quantity (\ref{eav}) can strongly depend on the decomposition.
Hence convex roof entanglement measures are defined as the average of pure state entanglement
minimized over all possible preparations of the state (decompositions),
\begin{equation}
\label{e}
E(\hat{\rho})=\min_{\alpha}\tilde{E}(\hat{\rho}).
\end{equation}

\section{The proof \label{sec4}}

We prove that eq.~(\ref{eog}) for all density matrices of the form (\ref{rhon})
with eq.~(\ref{rhonn}) fulfilled in three stages. We first study a QS composed of a single
qubit and an environment with only two subspaces. In Subsection \ref{sec4b} we 
generalize the results to QSs of any size and finally in Subsection \ref{sec4c}
we perform the simple generalization to environments with any number of subspaces.

\subsection{Qubit and environment with two subspaces \label{sec4a}}

Here we study the density matrix of a qubit
(the smallest possible QS) and an environment, which has two blocks with respect
to E. The disjoint subspaces of E are labeled as $E_1$ and $E_2$
and their subspaces are of dimension $N_1$ and $N_2$, respectively, with
$N_1+N_2=N$, where $N$ is the dimension of the whole E.
The density matrix can therefore be written as 
\begin{equation}
\label{mac}
\hat{\rho}=p_{1}\hat{\rho}_{1}+p_{2}\hat{\rho}_{2},
\end{equation}
where $p_{i}$ are probabilities with $p_{1}+p_{2}=1$
and $\hat{\rho}_{i}$ are density matrices of nontrivial dimension $2N_i$.

We assume that the pure state decomposition of each block 
that minimizes entanglement is known and we label the states by
$|\phi_i^k\rangle$, where $i=1,2$ distinguishes between the blocks,
so that 
\begin{equation}
\label{maci}
\hat{\rho}_{i}=\sum_k q_i^k|\phi_i^k\rangle\langle\phi_i^k|,
\end{equation}
 where $q_i^k$ are probabilities, with $\sum_k q_i^k=1$.
 Hence, entanglement of a given block is given by
 \begin{equation}
 \label{ent_blok}
 E(\hat{\rho}_{i})=\sum_k q_i^kE(|\phi_i^k\rangle).
 \end{equation}
 We will show that there does not exist any pure state decomposition of $\hat{\rho}$ for which
 entanglement is smaller than the weighed average of the entanglement present in each block,
 so
 \begin{equation}
 \label{ent}
 E(\hat{\rho})=p_{1}E(\hat{\rho}_{1})+p_{2}E(\hat{\rho}_{2}).
 \end{equation}
 
 To this end we will first study entanglement of any pure state,
 here written as a superposition of states which belong to the distinct blocks 
 \begin{equation}
 \label{stan}
|\psi\rangle = \alpha|\psi_1\rangle+\beta|\psi_2\rangle.
 \end{equation}
 These states can in turn be written as
 \begin{equation}
 \label{stani}
 |\psi_i\rangle=x_i|0\rangle\otimes|\varphi_i^0\rangle+y_i|1\rangle\otimes|\varphi_i^1\rangle.
 \end{equation}
 Environmental states $|\varphi_i^0\rangle$ and $|\varphi_i^1\rangle$
 do not have to be orthogonal when they belong to the same block,
 $\langle\varphi_i^0|\varphi_i^1\rangle \neq 0$, $i=1,2$,
 but they must be mutually orthogonal when they belong to different blocks,
 $\langle\varphi_1^a|\varphi_2^b\rangle = 0$, $a,b=0,1$.
 
 Hence, if we rewrite the state (\ref{stan}) in a form which simplifies the calculation
 of the reduced density matrix with respect to the qubit,
 \begin{equation}
 \label{stan2}
 |\psi\rangle = a|0\rangle\otimes|\psi_a\rangle+b|1\rangle\otimes|\psi_b\rangle,
 \end{equation}
 the parameters are given by 
 \begin{subequations}
 \begin{eqnarray}
 a&=&\sqrt{|\alpha|^2 |x_1|^2+|\beta|^2 |x_2|^2},\\
b&=&\sqrt{|\alpha|^2 |y_1|^2+|\beta|^2 |y_2|^2},
\end{eqnarray}
\end{subequations}
 while the normalized environmental states are 
 \begin{subequations}
 \begin{eqnarray}
 |\psi_a\rangle&=&\frac{1}{a}\left(\alpha x_1|\varphi_1^0\rangle
 +\beta x_2|\varphi_2^0\rangle\right),\\
 |\psi_a\rangle&=&\frac{1}{b}\left(\alpha y_1|\varphi_1^1\rangle
 +\beta y_2|\varphi_2^1\rangle\right).
 \end{eqnarray}
\end{subequations}
The reduced density matrix of E is then given by 
\begin{equation}
\hat{\rho}_E=a^2|\psi_a\rangle\langle\psi_a|+
b^2|\psi_b\rangle\langle\psi_b|,
\end{equation}
and the normalized linear entropy of the reduced density matrix,
which we will use as our pure state entanglement measure, 
is given by
\begin{equation}
\label{epsi}
E(|\psi\rangle)=2\left(1-\tr\hat{\rho}_E^2\right)=4a^2b^2\left(1-\left|\langle\psi_a|\psi_b\rangle\right|^2\right).
\end{equation}
Since the states of E from different subspaces have to be orthogonal to each other,
we obtain a simplified formula for the scalar product
\begin{equation}
\left|\langle\psi_a|\psi_b\rangle\right|^2
=\frac{1}{a^2b^2}\left|
|\alpha|^2x_1^*y_1\langle\varphi_1^0|\varphi_1^1\rangle
+
|\beta|^2x_2^*y_2\langle\varphi_2^0|\varphi_2^1\rangle
\right|^2.
\end{equation}

We will now compare the result with the average of the (normalized) linear entropy
of the block diagonal density matrix
\begin{equation}
\label{bd}
\hat{\rho}_{BD}=|\alpha|^2|\psi_1\rangle\langle\psi_1|
+|\beta|^2|\psi_2\rangle\langle\psi_2|;
\end{equation}
the matrix is a counterpart to the pure state (\ref{stan}), but without the inter-block coherences.
This is given by (the ``tilde'' signifies that this quantity is not in fact an entanglement
measure, since it has not been minimized)
\begin{equation}
\label{erho}
\tilde{E}(\hat{\rho}_{BD})=|\alpha|^2E(|\psi_1\rangle)+|\beta|^2E(|\psi_2\rangle),
\end{equation}
with
\begin{equation}
E(|\psi_i\rangle)=4|x_i|^2|y_i|^2\left(1-\left|\langle\varphi_i^0|\varphi_i^1\rangle\right|^2\right).
\end{equation}

The difference between the pure state entanglement given by eq.~(\ref{epsi})
and the average (\ref{erho}) is given by
\begin{eqnarray}
\label{roznica}
E(|\psi\rangle)-\tilde{E}(\hat{\rho}_{BD})&=&4|\alpha|^2|\beta|^2 \left[ \left(|x_1|^2 - |x_2|^2\right)^2 \right.\\
\nonumber
&&+ \left.\left| x_1^*y_1\langle\varphi_1^0|\varphi_1^1\rangle - x_2^*y_2\langle\varphi_2^0|\varphi_2^1\rangle\right |^2 \right]\geq 0.
\end{eqnarray}
This quantity is obviously always greater or equal to zero
which means that the entanglement present in a superposition of states from the different blocks
is always greater than the average of the entanglement  
contained in each block separately. Note that this result is only true because
the different blocks are disjoint in terms of the states of E.

Given the result (\ref{roznica}) it is now straightforward to show that 
entanglement present in a density matrix of the form (\ref{mac}) is found using eq.~(\ref{ent}).
Let us start with an arbitrary pure state decomposition of $\hat{\rho}$, eq.~(\ref{rhodec}),
which we henceforth label with the index $A$.
If we rewrite each state $|\psi_n\rangle$ as a superposition of states which belong to the distinct blocks 
as in eq.~(\ref{stan}) with coefficients $\alpha_n$ and $\beta_n$, we get
\begin{eqnarray}
\label{blop}
\hat{\rho}&=&\sum_n P_n|\alpha_n|^2|\psi_{1n}\rangle\langle\psi_{1n}|+
\sum_n P_n|\beta_n|^2|\psi_{2n}\rangle\langle\psi_{2n}|\\
\nonumber
&&+
\sum_n P_n\alpha_n\beta_n^*|\psi_{1n}\rangle\langle\psi_{2n}|+
\sum_n P_n\alpha_n^*\beta_n|\psi_{2n}\rangle\langle\psi_{1n}|.
\end{eqnarray}
Since the density matrix $\hat{\rho}$ is block diagonal with respect to the different subspaces,
the last two terms must be equal to zero, so
\begin{equation}
\label{prostszy}
\hat{\rho}=\hat{\rho}_B=\sum_n P_n|\alpha_n|^2|\psi_{1n}\rangle\langle\psi_{1n}|+
\sum_n P_n|\beta_n|^2|\psi_{2n}\rangle\langle\psi_{2n}|
\end{equation}
is a different pure state decomposition of the same density matrix,
we label this decomposition by $B$. The crucial difference between the two decompositions is that the latter
does not contain any states which encompass both subspaces, so we can write the density matrix
directly in the form given by eq.~(\ref{mac}) with individual decompositions,
\begin{equation}
\hat{\rho}_i=\frac{1}{p_i}\sum_n q_n^i|\psi_{in}\rangle\langle\psi_{in}|,
\end{equation}
with $q_n^1=P_n|\alpha_n|^2$, $q_n^2=P_n|\beta_n|^2$, and $p_i=\sum_n q_n^i$.
These are not necessarily the decompositions which minimize entanglement 
in each block separately (\ref{maci}).

Using the property (\ref{roznica}), we can show that the average of the (normalized) linear entropy
for decomposition $A$ is always greater or equal to the average for decomposition $B$,
since
\begin{eqnarray}
\tilde{E}(\hat{\rho}_A)&=&\sum_n P_nE(|\psi_n\rangle)\\
\nonumber
&\ge&
\sum_n P_n|\alpha_n|^2E(|\psi_{1n}\rangle)+
\sum_n P_n|\beta_n|^2E(|\psi_{2n}\rangle)\\
\nonumber
&=&\tilde{E}(\hat{\rho}_B).
\end{eqnarray}
Hence, for every pure state decomposition of the block diagonal density matrix $\hat{\rho}$, there exists a decomposition which does not contain superposition states between
the two subspaces of E, for which the average of pure-state entanglement is lesser or equal.
The direct consequence of this is that minimization of average entanglement over all possible 
pure state decompositions will yield the average of entanglement minimized in each block separately,
and mixed state entanglement is given by eq.~(\ref{ent}).

\subsection{Larger quantum system \label{sec4b}}
The generalization of the result of the previous section to a larger QS is rather straightforward.
The first step is to show that pure state entanglement is always larger than the average entanglement
of the counterpart block-diagonal density matrix, meaning that the inequality
\begin{equation}
\label{new}
E(|\psi\rangle)-\tilde{E}(\hat{\rho}_{BD})\ge 0
\end{equation}
still holds. Since we are still studying the situation with two subspaces, 
the relevant states are given by eqs (\ref{stan}) and (\ref{bd}), with the average linear entropy
of state $\hat{\rho}_{BD}$ definedy by eq.~(\ref{erho}).
We show that the inequality holds in Appendix \ref{apa}. 

Once this is established, we can show that for any decomposition ($A$) of the density matrix (\ref{rhodec})
there exists a counter decomposition ($B$) which only contains states limited to either block,
for which the average pure-state entanglement is smaller or equal. This is shown as in the previous section,
by writing each state in decomposition $A$ as in eq.~(\ref{stan})
to obtain eq.~(\ref{blop}), and then by noting that the off-diagonal terms that connect the two subspaces
must sum to zero, to obtain decomposition $B$, given by eq.~(\ref{prostszy}).
As the averaged entanglement in the two decompositions can, as previously, be connected using
the pure-state inequality (\ref{new}) yielding $\tilde{E}(\hat{\rho}_A)\ge\tilde{E}(\hat{\rho}_B)$,
it follows that any convex-roof entanglement measure of a density matrix which is block diagonal 
in two blocks which span different environmental subspaces is given by the average entanglement
contained in each block (\ref{ent}) regardless of QS size.

\subsection{More subspaces \label{sec4c}}
The generalization to more subspaces is even more straightforward. Obviously, since entanglement
can be found by averaging the entanglement between two subspaces, if there are $M$
block diagonal subspaces in terms of E states, the entanglement of the whole density matrix
can be found by averaging entanglement between one subspace and the remaining $M-1$,
while the entanglement of the $M-1$ blocks can be found by averaging between one of them
and the remaining $M-2$, and so on. Consequently the entanglement in a matrix of $M$ blocks
can be found by averaging over the entanglement of each block, and hence is given by eq.~(\ref{e}).

\section{Consequences \label{sec5}}

In this section we discuss the consequences of the central result of this paper, 
namely that any convex-roof entanglement measure can be calculated using eq.~(\ref{eog})
for density matrices which are block diagonal in disjoint subspaces, for the theory of mixed state
entanglement. Outside the obvious simplification of calculation of such entanglement measures,
since now minimization has to be performed separately over parts of the QS-E Hilbert space,
there are two significant implications.

The first pertains to bound entanglement. Negativity \cite{vidal02,lee00a,plenio05b}), the only measure to quantify
mixed state entanglement for larger systems
which can be found directly from the density matrix, is defined as the absolute sum of the negative eigenvalues 
of the density matrix after partial transposition with respect of one of the subsystems is performed.
In other words, it is a measure based on the Peres-Horodecki criterion \cite{peres96a,horodecki96},
which states that if the matrix after partial transposition is not a density matrix (has negative
eigenvalues) then there is entanglement in the state. The drawback of the measure and the criterion
is that they do not detect certain entangled states; such states are said to contain bound entanglement \cite{horodecki97b,horodecki98}. The existence of bound entanglement and its experimental demonstration
have received a lot of attention \cite{smolin01,diguglielmo11,hiesmayr13,sentis18,gabodulin19}, but the exact limitations on when bound entanglement can be expected 
are not known with the exception of small systems and states with additional symmetries. 

For density matrices which are block diagonal with respect to disjoint subspaces of E,
Negativity can obviously be found in each block separately,
since partial transposition of the density matrix with respect to E
does not mix the blocks and both the partial transposition and the calculation of eigenstates
can be done in each block separately. The question remains, if there exist bound entangled states
that transgress these blocks. Since we have shown that 
this property translates
to all convex-roof entanglement measures, we have automatically shown that bound entanglement cannot
exist between separate blocks of this type. It can only exist within a single block, hence imposing a
new limitation on states where we can expect this type of quantum correlations.

Practically this extends the set of states for which bound entanglement is impossible to systems of 
any size as long as individual disjoint blocks are not greater than $6\times 6$ matrices.
Less trivially, it also pushes the boundary for the rank of the studied bipartite state
of dimension $M\times N$ required for bound entanglement \cite{kraus00,horodecki00},
since each block can be treated separately.
Hence if the dimension of QS is $M$, while E is separated into $n$ disjoint blocks each of dimension
$N_i$, with $N=\sum_iN_i$, then the general criterion that guarantees only free entanglement
(not bound), namely that the rank of the density matrix has to be $\le\max\left(M,N\right)$
is supplemented by criteria for each block, which are easier to check because of the smaller size
of the blocks. For bound entanglement to be impossible in a given block, its rank must be
$\le\max\left(M,N_i\right)$.
Note that the block criteria do not always directly imply the criterion for the full density matrix,
since the dimension of a given block of the environment can be smaller than the dimension of QS. 

The second implication pertains to the existence of bipartite mixed states which
are classified by convex-roof entanglement measures as containing maximum entanglement.
Such states were considered unlikely by the original paper on the topic \cite{cavalcanti05},
but states of required form were found in Ref.~\cite{li12}. It has later been shown that the periodical
emergence of such states is possible in qubit-environment evolutions driven by Hamiltonians that lead
to pure dephasing of the qubit \cite{roszak20} and are important for the emergence of objectivity
\cite{ollivier04,ollivier05,zurek09,korbicz14,horodecki15,roszak19b,kwiatkowski21}.
Here we have found that the emergence of such states is not limited to pure-dephasing evolutions
and that they can manifest themselves in other classes of interactions.

\begin{figure*}[tb]
	\includegraphics[width=0.8\textwidth]{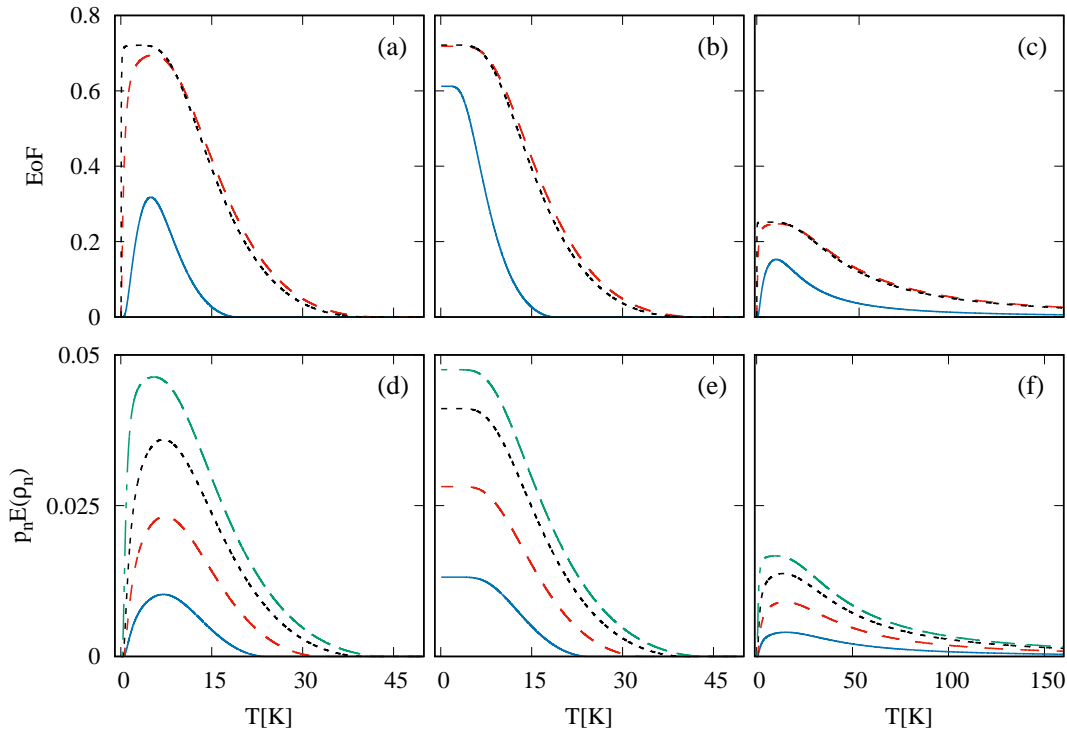}
	\caption{Entanglement at thermal equilibrium as a function of temperature for Hamiltonian
		formed from blocks given by eq.~(\ref{h2}). 
		Top panels show Entanglement of Formation for (a) finite eigenenergies $E_m$ and 
		non-zero magnetic field, $\Omega=1$, (b) finite eigenenergies $E_m$ and 
		zero magnetic field, $\Omega=0$,  (c) infinite eigenenergies $E_m=\infty$ and 
		non-zero magnetic field, $\Omega=1$. The three curves correspond to different sizes of the
		environment: $K = 1$ - solid blue line, $K = 10$ - dashed red line, $K = 100$ - dotted black line. Bottom panels show corresponding [$(a)\rightarrow (d)$, etc.] components of entanglement for different blocks, 
		$p_n E(\rho_n)$ for $K = 10$ with $m = -10$ - solid blue line, $m = -8$ - dashed red line , $m = -5$ - dotted black line, $m = 0$ - dashed green line.  }  \label{fig1}
\end{figure*}

\section{Relevance \label{sec6}}
The relevance of the presented proof is based on the answer to the question, if there exist
physical QS-E states which have the form (\ref{rhon}). The answer is yes and we will discuss
some situations when this is the case with special focus on thermal QS-E states.
In fact the requirement for a density matrix describing a state at thermal equilibrium corresponding 
to temperature $T$, $\hat{\rho}_T=\exp(-\beta \hat{H})/Z$, with $\beta=1/k_BT$ and $Z$ being the partition function, is that the Hamiltonian has the required form.

We will be taking into account the standard form of a Hamiltonian describing two interacting systems, so
$\hat{H}=\hat{H}_{\mathrm{QS}}+\hat{H}_{\mathrm{E}}+\hat{H}_{\mathrm{QS-E}}$, where the first term
describes the free Hamiltonian of QS, the second of E, and the third term describes their interaction.
We require the full Hamiltonian to be block diagonal in terms of the states of E. The form of the QS
part is therefore arbitrary, yet the interplay of the interaction and
free Hamiltonian of E is cruitial.

There is a plethora of systems (which would here constitute E) for which the Hamiltonian commutes
with at least one non-trivial observable $\hat{A}$, $[\hat{H}_E,\hat{A}]=0$. In nontrivial situations
(when the eigenstates of the observable $\hat{A}$ are degenerate in the full Hilbert space of
$\hat{H}_E$, or, physically speaking, the observable describes only one degree of freedom, such as spin,
of a more complex system) this yields a Hamiltonian which is block-diagonal in subspaces
corresponding to a single eigenvalue of $\hat{A}$.

A good example here is the Hubbard model \cite{giamarchi04,hubbard63,rasetti91,jafari08,baeriswyl13}
as it commutes with three different operators, each reducing the dimension of the blocks.
Firstly, the Hubbard Hamiltonian commutes with the number operator, meaning that each subspace
of the Hilbert space where the states describe an equal number of particles constitutes
a separate block. Furthermore, the Hamiltonian commutes with the total spin $\hat{S}_z$
component, so within each block there is a smaller block-diagonal structure,
which differentiates between states with different spin symmetry. Finally, it also commutes with the
parity operator, yielding even smaller non-trivial blocks. Such symmetries are common
in many-body systems and are seen in e.~g.~ the Heisenberg, Bose-Hubbard, t-J, or Holstain models \cite{mahan00,giamarchi04,gersch63,dagotto89,sowinski12}.

Obviously, the block-diagonal form of $\hat{H}_E$ is not sufficient for the full Hamiltonian
to demonstrate block-diagonality with respect to environmental subspaces. To this end the
interaction term must possess the same symmetries as those that yield the structure of $\hat{H}_E$.
Namely if the Hamiltonian of E commutes with the observable $\hat{A}$, then also the interaction
term must commute with it, $[\hat{H}_{int},\hat{A}]=0$.
This means that during the interaction the quantum system would be susceptible to the effect of the
environment with respect to some degree of freedom
while another (described by the observable $\hat{A}$) would not affect it.

If the Hamiltonian has the specific block-diagonal structure not only thermal equilibrium states
will retain it, but it will be kept during time-evolution as long as the initial QS-E state 
is a mixture of states contained within single blocks. The most natural situation when this is obtained
is when the initial QS-E state is of product form
with the initial E state being a thermal equilibrium state of block-diagonal $\hat{H}_E$.
There are then no limitations on the initial states of QS.

The initial QS-E state can also be used to guarantee that the block-diagonal form of the
full density matrix with disjoint subspaces is kept throughout an evolution governed 
by a Hamiltonian where the subspaces of the blocks partially overlap.
An example of such a (full) Hamiltonian is one describing a spin interacting with an environment
of spins via the hyperfine interaction
in the ``box model'' approximation \cite{merkulov02,melikidze04,barnes11,shulilin21}. 
Such a Hamiltonian is block-diagonal with disjoint subspaces with respect
to the total spin operator of E, but the block-diagonality which is present in subspaces
governed by the projection of the total spin operator overlap.
By choosing an initial environmental state
which is a mixture of states contained within these smaller subspaces, but skipping some spin quantum
numbers, one could guarantee that the QS-E density matrix would have the required form
throughout the evolution.

\section{Examples \label{sec7}}

As an example we will study a Hamiltonian describing a qubit interacting with an environment,
where the Hamiltonian is block-diagonal in terms of $4\times 4$ blocks,
where each block is composed of the same $|\!\uparrow\rangle,|\!\downarrow\rangle$
qubit states
and of a different set of E states, $|m_{1}\rangle$ and $|m_{2}\rangle$.
Hence each block is given in the QS-E basis
$\{|\!\uparrow m_{1}\rangle,|\!\uparrow m_{2}\rangle,
|\!\downarrow m_{1}\rangle,|\!\downarrow m_{2}\rangle\}$
and the different blocks are distinguished by ``quantum number'' $m$.
We limit the size of the blocks, so that we can use Wootters formula for two-qubits
to quantify Entanglement of Formation \cite{wootters98} in each block,
while studying the different possible thermal behaviors of the entanglement of the whole
state.

We study a Hamiltonian that is composed of blocks given by
\begin{equation}
\label{h2}
\hat{H}^m=\left(
\begin{array}{cccc}
E_{m1}&0&0&M_m\\
0&E_m&0&0\\
0&0&E_m&0\\
M_m^*&0&0&E_{m2}
\end{array}
\right),
\end{equation}
so the full Hamiltonian is $\hat{H}=\sum_m \hat{H}^m$.
The blocks are intentionally chosen in such a way that they have two entangled eigenstates
and two separable ones. 
The Hamiltonian parameters are
\begin{subequations}
	\label{par}
	\begin{eqnarray}
	E_{m1}&=&\alpha \left(m + \frac{\Omega}{2} \right),\\
	E_{m2}&=&-\alpha \left(m + 1 + \frac{\Omega}{2}\right),\\
	M_m&=&\alpha\sqrt{K(K+1)-m(m+1)},
	\end{eqnarray}
\end{subequations}
mimicking spin systems, with $m=0,\pm 1, \pm 2,\dots,\pm K$, where $K=\max(m)$ is an integer.
The parameter $\alpha=1/\sqrt{K}$ eV is responsible for the strength of the QS-E interaction.
The scaling with $\sqrt{K}$ is responsible for the interaction with the whole environment
to be equivalent regardless of $K$ for large values of $K$, which is in accordance with the scaling prevalent for quantum dot spin qubits \cite{barnes11,mazurek14b}.  
The parameter $\Omega$ plays the role of the magnetic field.
$E_m$ denote the energies of the separable eigenstates.

In Fig.~\ref{fig1} in the top panels we plot EoF between the qubit and the environment at thermal equilibrium
as a function of temperature for the Hamiltonian.
The figures correspond to three sets of Hamiltonian parameters which lead to qualitatively different
behavior of entanglement. Plot (a) and (b) have separable eigenenergy
specified as $E_m=\frac{1}{2}\left(E_{m_1}+E_{m_2}\right)$, meaning that within a single block
the entangled eigenstates constitute the ground state and the state with the highest energy.
For Fig.~\ref{fig1} (a) the magnetic-field parameter
$\Omega = 1$ is nonzero lifting the degeneracy between the blocks, while for (b) $\Omega = 0$
which leads to degenerate eigenstates (all four eigenstates within a given block have no $m$ dependence). In Fig.~\ref{fig1} (c) we have $\Omega = 1$ as in (a), but the
separable eigenstates are much higher in energy, $E_m=\infty$, which means that the Hamiltonian
effectively has two-dimensional blocks.
The three curves on each plot correspond to $K=1$ with $3$ blocks (blue, solid line), $K=10$
with $21$ blocks (dashed, red line),
and $K=100$ with $201$ blocks (dotted, black line).

Fig.~\ref{fig1} (a) shows a fast growth of entanglement followed by slower decay which is cut short
by ``sudden death'' \cite{rajagopal01,horodecki01b,yu04,yu09} and beyond a certain temperature there is no entanglement.
The rate of the growth strongly depends on the number of blocks, so the difference between the 
$K=10$ and $K=100$ curves is obvious only on the low-temperature side.
There is no entanglement at zero-temperature; this is easy to understand, since
the ground state of the whole Hamiltonian is the lowest energy state of the $m=K$ block, which is 
separable, since $M_K=0$. Fig.~\ref{fig1} (d) contains the temperature-dependence of 
a choice of components which correspond to different blocks in the whole EoF curve for $K=10$
[the probability of a given block times EoF within the block as in eq.~(\ref{eog})].
It is interesting to note that ``sudden death'' occurs at different temperatures for different
components which should yield points in the corresponding full entanglement curve which
are not smooth. These points are not visible because the ``sudden death'' resulting
from the rising mixedness of the Gibbs state with temperature occurs much more gently
(although undeniably it is present) than seen 
typically in evolution \cite{yu04,roszak06b,muzzama18,wang18,subhadeep19}.

Fig.~\ref{fig1} (b) and (e) correspond to a situation where there is a degenerate ground state
showing entanglement of the full state and its components, respectively.
The lowest-energy state of each block is also the ground state of the Hamiltonian; since these states 
are entangled (with the exception of the $m=K$ state), the zero-temperature state
is a mixture of entangled states from different blocks and is itself entangled.
With rising temperature we only observe a decay of entanglement corresponding to 
the state of individual block becoming mixed.
The decay is again not exponential due to ``sudden death'' which occurs at some finite temperature.

Note that the striking qualitative difference between the temperature-dependencies of entanglement
in Fig.~\ref{fig1} (a) and (b) are not a result of different entanglement behavior within the blocks.
In fact the blocks in both cases behave qualitatively the same, but the probabilities with which
the entanglement of a given block contributes to overall entanglement is very different in the two cases.
The results of Fig.~\ref{fig1} (a) show a trade off between the rising probability of a given block
to occur within the Gibbs state versus entanglement becoming smaller when the state becomes more mixed,
while in the results of Fig.~\ref{fig1} (b) the diminishing of entanglement closely follows the
decay of entanglement within the blocks.

This changes in Fig.~\ref{fig1} (c) and (f) where the blocks of the Hamiltonian are effectively
two-dimensional which leads to ``sudden death'' type behavior being impossible due to the
geometry of separable states \cite{zyczkowski98,bengtsson06,roszak06b}. The situation is equivalent
to that of Fig.~\ref{fig1} (a) with the exception that only two eigenstates within each block, the
entangled ones, are parts of the Gibbs state. This leads to no ``sudden death'' in the components
and therefore no ``sudden death'' in the entanglement of the full state, so actual separability
is reached only at infinite temperature. There is also an obvious effect visible in the 
maximum entanglement which can be present at thermal equilibrium, which is much smaller.
This is because mixing of two orthogonal entangled states which are confined to the same two-dimensional
subspace is much more detrimental to entanglement than the admixture of separable states.
Hence here, although the probabilities with which the blocks are mixed behave very similarly
to that in Fig.~\ref{fig1} (a), it is the different temperature-dependence in individual blocks
that leads to qualitatively different results.

\begin{figure}[tb]
    \centering
    \includegraphics{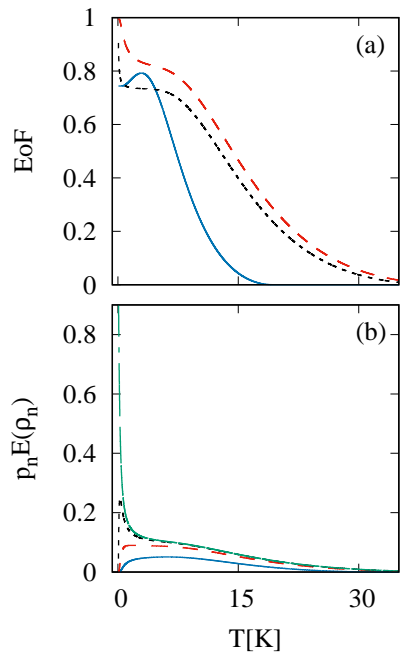}
    \caption{Entanglement of Formation as a function of temperature T for
    	finite eigenenergies $E_m$ and 
    	non-zero magnetic field, $\Omega=1$, considering only non-positive values of $m$.
    Top and bottom panel as in Fig.~\ref{fig1}  }
    \label{fig2}
\end{figure}

In Fig.~\ref{fig2} we show the same situation as in Fig.~\ref{fig1} (a), but where only blocks with
non-positive values of $m$ have been taken into account. This eliminates the block with $m=K$
which has only separable eigenstates and also contains the ground state of the whole Hamiltonian.
Now the ground state is also entangled which leads to a more complicated trade-off situation
between the components shown in the bottom panel of the figure starting with an entangled
Gibbs state at zero-temperature. The corresponding component of entanglement decays very rapidly with temperature,
while all other components start to grow which leads to a plateau at small temperatures for
larger $K$ and an initial growth of entanglement for $K=1$.

We have shown that the temperature-dependence of entanglement can be very diverse
even in case of the simple Hamiltonian which has been studied. This fact is obvious in terms
of possible behavior within a given block of the density matrix, but our study shows
that the interplay of probabilities pertaining to each block can be just as important.
By changing the way that temperature mixes the different blocks in the Gibbs state,
we have obtained three qualitatively different scenarios, varying not only in zero-temperature
entanglement, but also in monotonicity for different temperature regimes. It is important to note
that the curves were obtained by qualitatively the same Hamiltonian, and a larger change
was neccesary only to modify the behavior of entanglement within a single block.

\section{Conclusion \label{sec8}}

We have supplied a proof that any convex-roof entanglement measure is an average 
over entanglement within a given block of a density matrix which is block diagonal 
in such a way that individual blocks are contained within different subspaces 
of one of two potentially entangled subsystems
(say, the environment). Although this may seem to be an overly specific regime of application,
there are many Hamiltonians which possess this quality. For such Hamiltonians, any
thermal-equilibrium state will inherit the same quality and the simplification of resulting
calculation of entanglement can be immense. If an initial bipartite state is also block
diagonal in such a way, then the evolution driven by the Hamiltonian will yield 
a state which retains the necessary form at any given time. 

We have used the method to find the temperature-dependence of entanglement in a Gibbs state
of a large system, the Hamiltonian of which is composed of a multitude of small blocks.
This allowed us to show four very different 
behaviors of entanglement which can be occur in different parameter ranges. 
This
can be dependent on how entanglement reacts to temperature change within each block,
but it also strongly depends on the way that different blocks are mixed. We have shown
that the interplay of probabilities can yield striking differences in the trends 
of entanglement, both in terms of entanglement at zero temperature,
as well as weather it is ascending or descending in a given temperature range. 
Temperature driven ``sudden death'' type behavior is the singular property which requires
``sudden death'' behavior to occur within each block separately.

On the more quantum-theoretical side, our proof allows us to infer that bound entanglement cannot
traverse multiple blocks; it is possible only within a single block. This has consequences for 
the calculation of such measures as Negativity, which cannot detect bound entanglement, making
them more reliable.

\begin{acknowledgements}
This work was supported by project 20-16577S of the Czech Science Foundation.
\end{acknowledgements}

\appendix
\section{Qudit and environment with two subspaces \label{apa}}
Here we will show that entanglement of any pure state that contains coherences between subspaces
of E, eq.~(\ref{stan}) is greater or equal to the average entanglement in its block-diagonal 
counterpart, eq.~(\ref{bd}) regardless of the size of QS.
Since we are now dealing with a qudit, a system of dimension $K$, the states (\ref{stani})
have to be substituted by 
\begin{equation}
|\psi_i\rangle=\sum_sx_i^s|s\rangle\otimes|\varphi_i^s\rangle,
\end{equation}
where the summation over $s$ spans the the qudit Hilbert space and the states $|s\rangle$
are a set of basis states on QS; $i=1,2$ still differentiates between the subspaces of E.
This yields the qudit version of eq.~(\ref{stan2}),
\begin{equation}
|\psi\rangle=\sum_sa_s|s\rangle\otimes|\psi_s\rangle,
\end{equation}
with
\begin{equation}
a_s=\sqrt{|\alpha|^2|x_1^s|^2+|\beta|^2|x_2^s|^2}
\end{equation}
and
\begin{equation}
|\psi_s\rangle=\frac{1}{a_s}\left(\alpha x_1^s|\varphi_1^s\rangle
+\beta x_2^s|\varphi_2^s\rangle\right).
\end{equation}

The reduced density matrix of E is given by 
\begin{equation}
\hat{\rho}_E=\sum_sa_s^2|\psi_s\rangle\langle\psi_s|
\end{equation}
and the normalized linear entropy of the reduced density matrix,
is 
\begin{equation}
E(|\psi\rangle)=4\sum_{s\neq s'}a_s^2a_{s'}^2\left(1-\left|\langle\psi_s|\psi_{s'}\rangle\right|^2\right).
\end{equation}
Since the states of E from different subspaces have to be orthogonal to each other,
the simplified formula for the scalar product is still valid and we have
\begin{equation}
\left|\langle\psi_s|\psi_{s'}\rangle\right|^2
=\frac{1}{a_s^2a_{s'}^2}\left|
|\alpha|^2x_1^{s*}x_1^{s'}\langle\varphi_1^s|\varphi_1^{s'}\rangle
+
|\beta|^2x_2^{s*}x_2^{s'}\langle\varphi_2^s|\varphi_2^{s'}\rangle
\right|^2.
\end{equation}

The average of the
linear entropy of the block diagonal density
matrix (\ref{bd}) is still given by eq.~(\ref{erho}), but with 
\begin{equation}
E(|\psi_i\rangle)=4\sum_{s\neq s'}|x_i^s|^2|x_i^{s'}|^2\left(1-\left|\langle\varphi_i^s|\varphi_i^{s'}\rangle\right|^2\right).
\end{equation}
Since the formulas in both cases only differ by an identical sum over different qudit basis states,
it follows that the formula (\ref{ent}) holds for a QS of any size.

\end{document}